\documentstyle[aps,preprint]{revtex}

\begin{document}
\begin{center}
{\bf Equation of state of finite nuclei and liquid-gas phase transition}\\
\vspace*{0.5 true in}
J. N. De\footnote{jadu@veccal.ernet.in}\\    

Variable Energy Cyclotron Centre, 1/AF, Bidhannagar, Calcutta 700 064, India\\
B. K. Agrawal\footnote{bijay@tnp.saha.ernet.in}
, S. K. Samaddar\footnote{samaddar@tnp.saha.ernet.in} \\
Saha Institute of Nuclear Physics, 1/AF, Bidhannagar, Calcutta 700 064,
India\\
\end{center}
\vspace*{.75 true in}
\begin{abstract}
 We construct the equation of state (EOS) of finite nuclei 
including surface and Coulomb effects in a
Thomas-Fermi framework using a finite range, momentum and
density dependent two-body interaction. We identify critical temperatures
for nuclei below which the EOS so constructed shows clear signals for
liquid-gas phase transition in these finite systems. 
Comparison with the EOS of infinite nuclear matter shows that the
critical density and temperature of the phase transition in nuclei
are influenced by the mentioned finite size effects.
\end{abstract}

\newpage

The equation of state of infinite nuclear matter calculated in the mean-field
approximation\cite{jaq,ban} shows a typical Van-der Waals behaviour. Below the
critical temperature $T_c$ ($\simeq 16\> MeV$), the liquid and the
gas phases are seen to coexist.  
A self-consistent determination of the EOS of a finite 
nuclear system  is however still not available. The nature of the 
phase transition or even its occurrence may depend on the
surface effects as well as on the Coulomb interaction between the protons.
From the experimental  side, the mass or charge distributions
from energetic proton\cite{fin} or heavy ion induced 
reactions\cite{lyn,chi}
show a power law behaviour \cite{fis} indicating a possible 
liquid-gas phase transition \cite{sie}
in finite nuclei. Recent experimental data on caloric
curve also shows strong signals for liquid-gas phase transition.
In the GSI data\cite{poc}  for $Au+Au$ at 600$AMeV$ the
temperature remains practically constant at $\simeq 5\> MeV$ in the 
excitation energy range of 3$-$10 $MeV$ per nucleon beyond which
the excitation energy increases linearly with temperature as in a
classical gas which is suggestive of a sharp phase transition. 
Analysis of the data from the EOS collaboration\cite{hau,ell}
for $Au+C$ at $1A\>GeV$ also indicates a liquid-gas phase transition
in finite nuclear systems.
It would therefore be of utmost 
importance to investigate the EOS of finite nuclei in a self-consistent
approach that would help in getting a clearer picture of the occurrence 
of phase transition in finite systems. 
An attempt in this direction is made in the Thomas-Fermi (TF)
framework\cite{de} in the present communication.

In the TF framework, the energy density of a nucleus of mass number
$A$ and proton number $Z$  is constructed from a Seyler-Blanchard
type\cite{sey} momentum  and density dependent effective 
interaction\cite{ban,mye} of finite range. The interaction is given by
\begin{equation}
v_{eff}(r,p,\rho)=C_{l,u} [v_1(r,p)+v_2(r,\rho)],
\end{equation}
\begin{equation}
v_1=-(1-\frac{p^2}{b^2})f(r),
\end{equation}
\begin{equation}
v_2=d^2[\rho(r_1)+\rho(r_2)]^n
\end{equation}
with 
\begin{equation}
f(r)=\frac{e^{-r/a}}{r/a}.
\end{equation}
Here $r= \mid {\bf r}_1-{\bf r}_2\mid$ and $p=\mid {\bf p}_1
-{\bf p}_2\mid$ are the separation of the two interacting nucleons in 
configuration and momentum space and $\rho(r_1)$ and $\rho(r_2)$ are the
densities at the sites of the two  nucleons. The quantities 
$C_{l}$ and $C_u$ are the strengths of the interaction between like
pair ($n-n$) or ($p-p$) and unlike pair ($n-p$), respectively. The values of the 
potential parameters determined from a fit of the well known bulk nuclear
properties are given  in Ref.\cite{de}. The incompressibility of
nuclear matter is then calculated to be 238 $MeV$. The Coulomb interaction
energy density is given by the sum of the direct and exchange 
terms\cite{de}. The energy density profile at a particular temperature $T$ 
is then given by 
\begin{equation}
\epsilon(r)= \sum_\tau \rho_\tau (r)\{ T J_{3/2}[\eta_\tau(r)]/
J_{1/2}[\eta_\tau(r)]
[(1-m^*_\tau(r) V_\tau ^1 (r)]+\frac{1}{2} V_\tau^0(r)\}
.
\end{equation}
Here $\tau$ is the isospin index, the $J$'s are the Fermi integrals
and $V_\tau^0$ is the single-particle potential which includes
the Coulomb term for protons. The momentum dependence in the interaction 
gives rise to $V_\tau^1$ which determines the effective nucleon
mass $m^*_\tau$. The fugacity $\eta_\tau(r)$ is defined
by 
\begin{equation}
\eta_\tau(r)= [\mu_\tau-V_\tau^0(r)-V_\tau^2(r)]/T,
\end{equation}
where $\mu_\tau$ is the chemical potential and $V_\tau^2$ is the
rearrangement energy term originating from the density
dependence in the interaction. The total energy per particle is then
given by
\begin{equation}
e(T)=\frac{1}{A} \int \epsilon(r) d{\bf r} .
\end{equation}
The entropy per particle, from the Landau quasi-particle approximation,
can be similarly calculated as 
\begin{equation}
s(T)=\frac{1}{A}\int  \sum_\tau \rho_\tau (r) 
\left [ \frac{\frac{5}{3} J_{3/2}[\eta_\tau(r)]}
{J_{1/2}[\eta_\tau(r)]}\right ].
d{\bf r}
\end{equation}
The free energy per particle is given by 
$f=e-Ts$. The density profiles of
neutrons and protons at different confining volumes are determined 
self-consistently for a fixed temperature. The pressure is then determined
from 
\begin{equation}
P=-\left ( \frac{\partial F}{\partial V}\right )_T,
\end{equation}
where $F=Af$ is the total free energy and $V$ stands for the confining
volume of the nucleus. The isotherms at different temperatures can then
be obtained. For the sake of comparison, the EOS of infinite
nuclear matter is also calculated, expressions for which are  
simpler and are given in Ref.\cite{ban}. 

The EOS of infinite symmetric nuclear matter is displayed
in the top  panel of Fig. 1. 
It resembles closely that for the Van-der Waals systems.
The isotherms are shown
for three temperatures, namely at $T=13.0 \> MeV$,
14.0 $MeV$  and   at the critical temperature $T_c$ which is 
found to be 14.5 $MeV$.  In the isotherms  the pressure
is plotted as a function of $V/V_0=\rho_0/\rho$ where $\rho_0$
(=0.153 $fm^{-3}$) is the saturation density at zero 
temperature. The dotted 
line refers to the liquid-gas coexistence curve and the dashed line 
represents the spinodal line. We also display 
in the bottom panel the EOS of asymmetric
nuclear matter with a representative asymmetry $X=(N-Z)/A=0.16$.
The medium heavy nuclei, namely, $^{85}Kr$ and
$^{150}Sm$ that we study in this communication have asymmetries close to 
0.16. The critical temperature for this asymmetric nuclear matter
decreases to $T_c=14.1$ $MeV$. The coexistence curve as well as the spinodal
line are also presented.

The isotherms for the nucleus $^{85}Kr$ are displayed
in Fig. 2 at four temperatures $T=$ 9.5, 10.5, 11.5 and
12.5 $MeV$ respectively. The critical temperature for this
system is found to be $T_c=11.5 \> MeV$. Here the abscissa is $V/V_0$ 
where $V_0$ is the  volume of the nucleus at zero temperature 
taken to be $V_0=\frac{4}{3}\pi r_0^3A$ with $r_0=1.16\>fm$ and
$V$ is the confining volume in which the self-consistent density profiles
are calculated. The isotherms for the finite nuclei are 
not much different in structure from those for infinite nuclear matter.
The liquid-gas 
coexistence curve can be drawn following the
Maxwell construction which is shown by the dotted line.
The spinodals are represented by the dashed line. The notable difference 
between the spinodal line  for infinite matter and that for the finite
system lies in the rising part AO (Fig. 2) 
which is slanted backwards for the latter. In Fig. 3, the isotherms for
$^{150}Sm$ are shown along with the coexistence curve and the spinodal line.
The critical temperature for this nucleus is $T_c$= 11.8 $MeV$. The
rising part AO of the spinodal line for this system 
is also a little backward-slanted.
The origin of this backward slant is traced back to the surface and Coulomb 
effects. We have checked that the slant changes from backward to forward 
with increasing size of the system. 

The critical parameters $T_c$, $P_c$ and the scaled critical
volume $V_c/V_0$ for symmetric and asymmetric ($X= 0.16$)
nuclear matter and of the finite nuclei $Kr$ and $Sm$ are listed
in Table 1. To see the role of asymmetry more clearly on the critical
parameters for infinite nuclear matter, the results with $X=0.5$
are also presented in the Table. 
With asymmetry, the critical temperature and pressure 
for the infinite system decreases while the critical volume increases.
For the finite systems, the critical temperature and pressure are considerably 
less than those for the infinite matter while the scaled critical volume
is significantly larger. The differences arise due to the surface 
and Coulomb effects. To delineate  the surface 
effects, we repeated the calculations for $Kr$ and $Sm$ switching off
the Coulomb interaction. From the Table, it is clear
that both these effects act in the same direction. However, the
mass region we consider shows the predominance of surface over the Coulomb 
effects. 

In order to highlight the difference in the coexistence curves
for finite nuclei and infinite nuclear matter,
we display in Fig. 4 the phase diagram
for infinite symmetric ($X$=0)  and
asymmetric ($X=0.5$) nuclear matter and for the nuclei $^{150}Sm$ and
$^{85}Kr$. The ordinate refers to the temperature scaled by
$T_c$, the critical temperature corresponding to each of the 
four systems mentioned and the abscissa refers to the scaled volume
$V/V_0$, obtained through Maxwell's construction from the respective
isotherms. The full and the dotted lines refer to the coexistence
curves for infinite matter with $X=0$ and 0.5, respectively; the
dashed and dash-dot curves correspond to $^{150}Sm$ and $^{85}Kr$.
The increase  in the critical volume with increasing asymmetry
and decreasing mass is clear. It may be mentioned 
that in the phase diagrams we have
displayed the scaled volume rather than the scaled density, as for a
finite nucleus, the density is not constant in contrast to infinite
nuclear matter. We may further mention that fluctuation effects due to
finite number of particles may smear the phase transition, particularly
near the critical temperature; blurring of the phase transition
due to the finite size has been discussed in the literature 
earlier\cite{imr,lab,goo}.

A discontinuity in the heat capacity $C_v$ for an infinite 
system or a bump in $C_v$ for a finite system\cite{hel} 
signals a phase transition.  Our calculation also shows a 
peak structure in $C_v$ (Fig. 5); the temperature at which 
peaking occurs depends on the choice of the confining volume.
With increasing volume (beyond $V_c$) the transition temperature
is smaller and the peak is sharper. We further see that for any
specified confining volume larger than the critical volume,
the transition temperature corresponds to that temperature
the isotherm for which shows a maximum at  the chosen volume.
It is therefore obvious that the incompressibility at the transition
temperature would vanish (or the compressibility would show a 
singularity ) which confirms further the occurrence 
of the phase transition. At this point on the isotherm, it is found
that the density profile self-consistently evolves to a nearly
uniform phase\cite{de2}, the so-called low density gas phase
that condenses out to the fragments of nuclei\cite{fis}.  
In an exact calculation, the transition
should occur at the crossing of the coexistence 
curve\cite{fis2} defined here by the Maxwell construction;
our calculation done in the mean-field approach
being an approximate one shows the transitions on the spinodal line
at volumes larger than the critical volume. The dependence
of the critical volume on the nuclear mass may have an important
bearing in understanding the nature of phase transition deduced 
from the shape of the fragment  mass or charge distributions
obtained in statistical models by fixing a 'freeze-out' volume
where the system is assumed to be homogeneous. The critical
volume possibly defines its lower bound. 

To summarize, we have calculated the equation of state of finite
nuclei in a self-consistent mean-field theory including surface and 
Coulomb effects. The critical parameters for finite nuclei, namely,
the critical temperature, pressure and volume are found to be
system dependent and differ significantly from those corresponding to
nuclear matter.  The 
nature of the isotherms,  the peaked structure of the heat
capacity and the singularity in the compressibility as obtained unambiguously
point to a liquid-gas phase transition in finite nuclear systems. 

\newpage
\noindent {\bf Figure Captions:}
\begin{itemize}
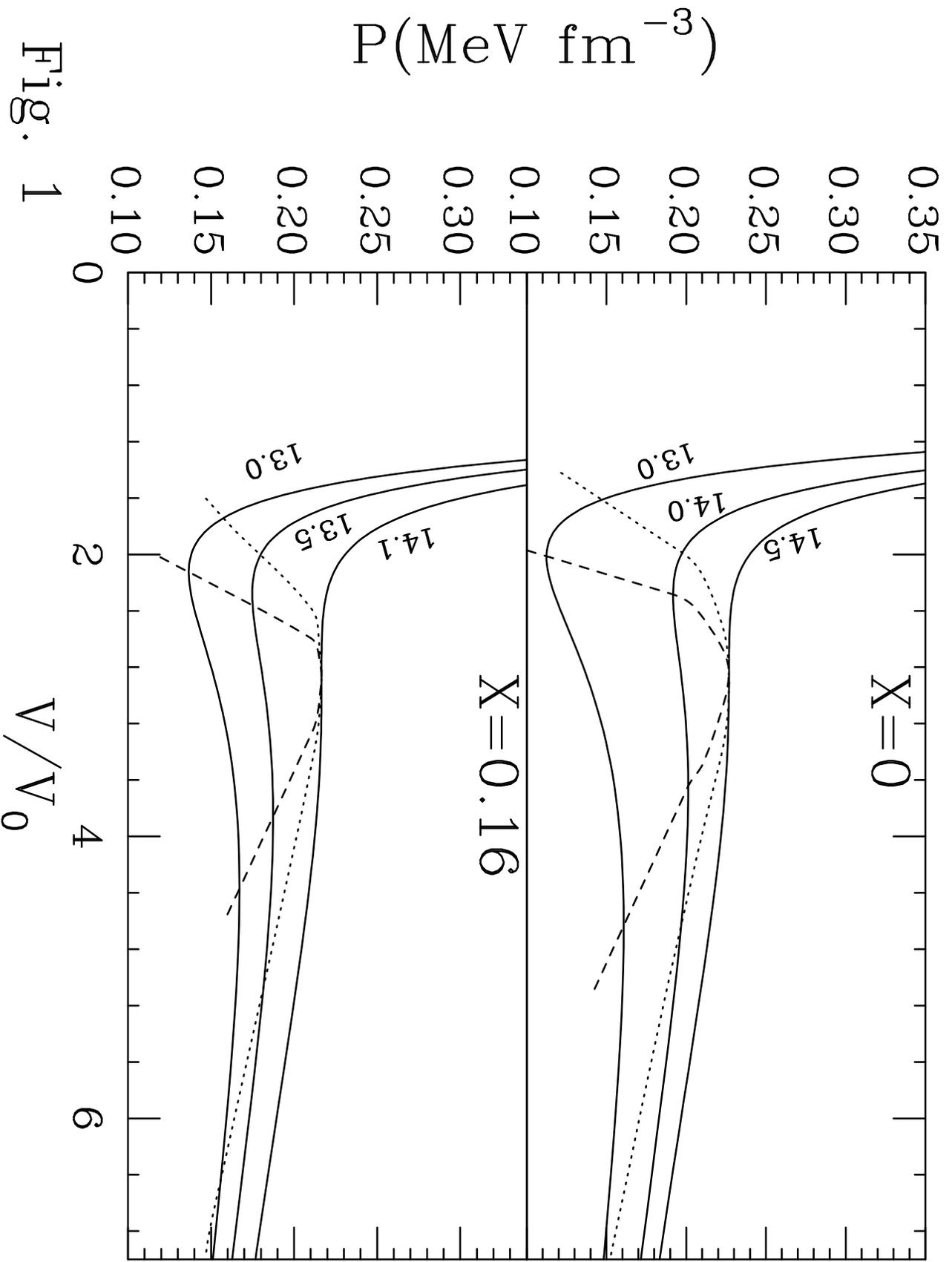
\item[Fig. 1] { The equation of state of symmetric nuclear matter (top panel)
and of asymmetric nuclear matter with asymmetry $X=0.16$ (bottom panel).
The temperatures (in $MeV$) for the isotherms are as marked in the figure.
The dotted lines are the coexistence curves and the dashed lines are the 
spinodals.
}
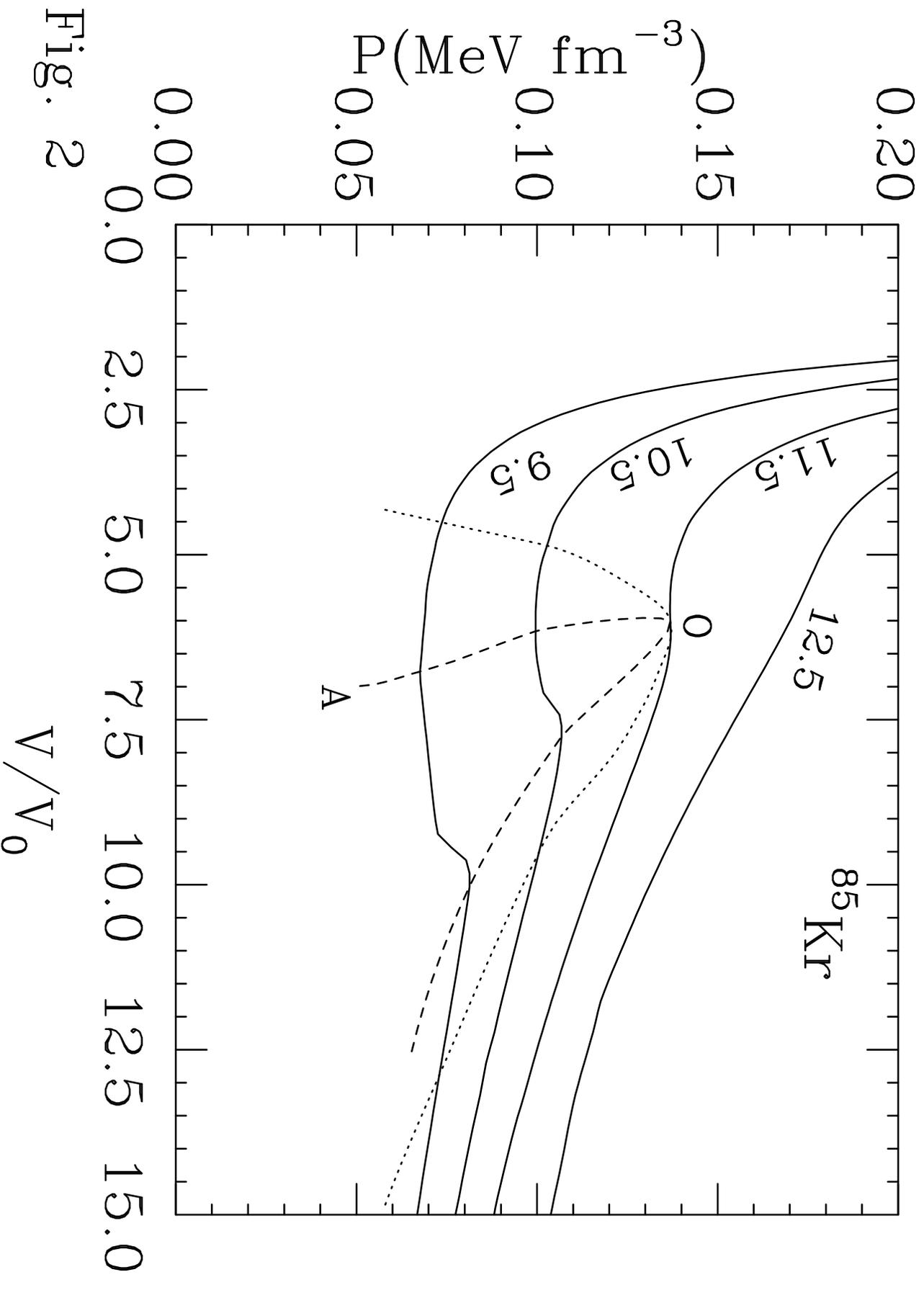
\item[Fig. 2] { The equation of state for the nucleus $^{85}Kr$. The
different notations used  have the same meaning as in Fig. 1. 
For further details see the text.}
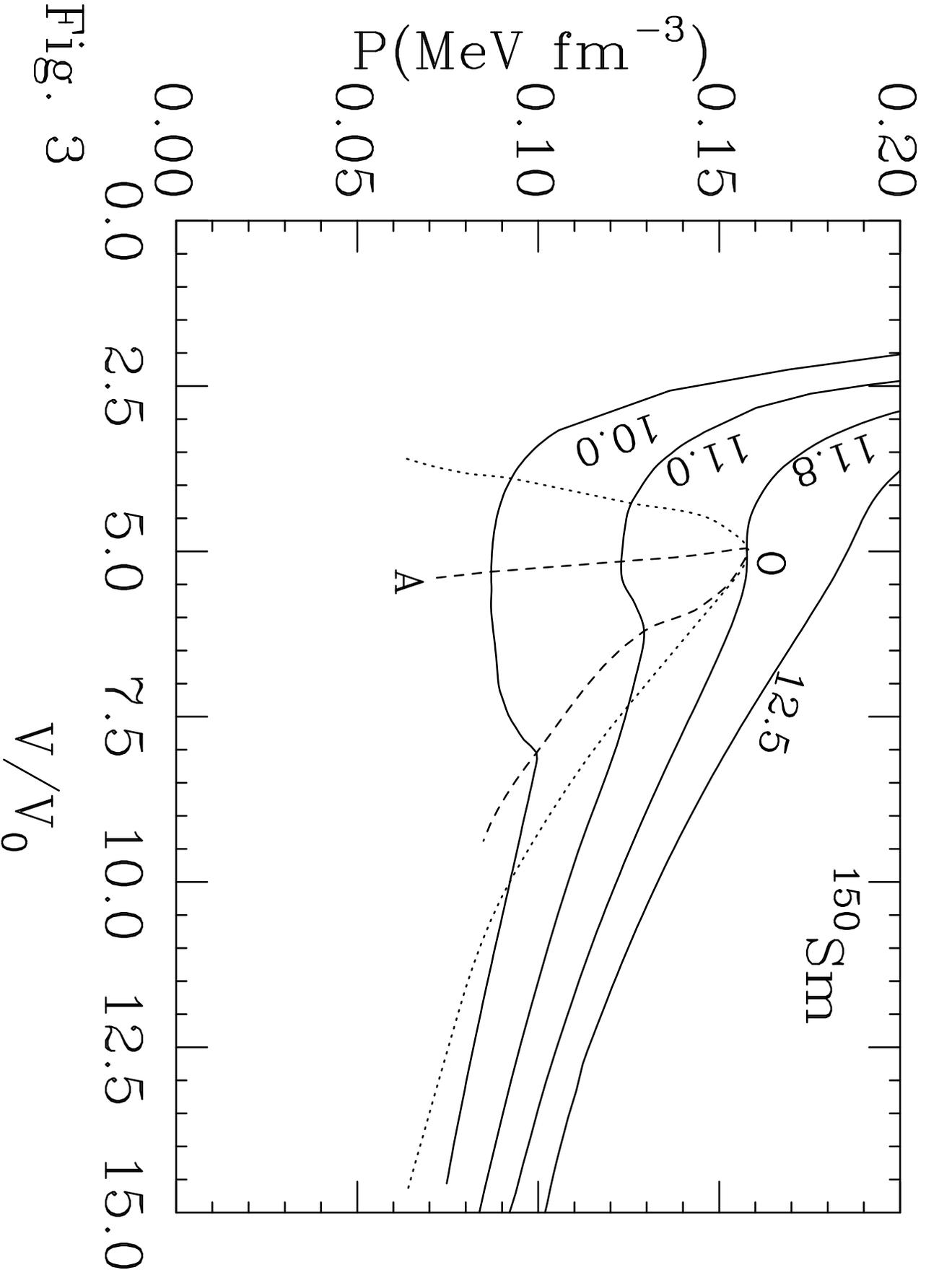
\item[Fig. 3] { Same as in Fig. 2 but for the nucleus $^{150}Sm$.}
\item[Fig. 4] { The phase diagram for symmetric and asymmetric nuclear matter
($X = 0$ and $X=0.5$ ) and for the nuclei $^{150}Sm$ and
$^{85}Kr$.} 
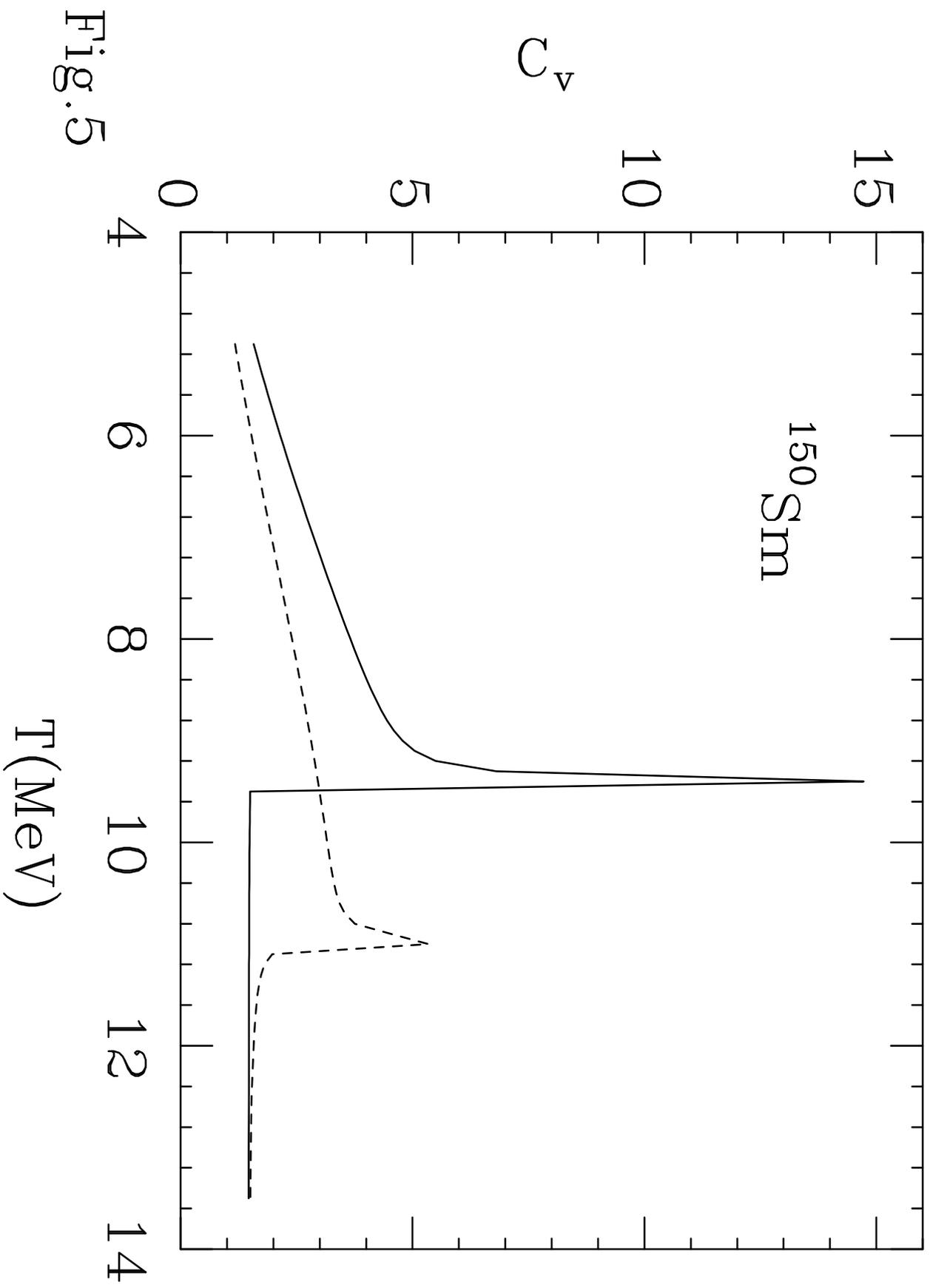
\item[Fig. 5] { The specific heat at constant volume $C_v$ plotted
as a function of temperature for the system $^{150}Sm$.
The full line and the dashed line correspond to the confining volumes
 $10V_0$ and $6V_0$, respectively.}
\end{itemize}
\newpage
\begin{table}
\caption{ Critical temperature, pressure and volume for a few systems}
\begin{tabular}{|cccc|}
\multicolumn{1}{|c}{System}&
\multicolumn{1}{c}{$T_c(MeV)$}&
\multicolumn{1}{c}{$P_c(MeV\> fm^{-3}$)}&
\multicolumn{1}{c|}{$V_c/V_0$}\\
\hline
Symmetric nm & 14.5& 0.227& 2.83\\
Asymmetric nm ($X=0.16$)& 14.1 & 0.216& 2.86\\
Asymmetric nm ($X=0.5$) & 10.5& 0.135& 3.40\\
$^{85}Kr$ (with Coulomb)& 11.5 & 0.137& 6.01\\
$^{85}Kr$ (no Coulomb)& 12.1 & 0.135& 6.66\\
$^{150}Sm$ (with Coulomb)& 11.8& 0.158& 4.95\\
$^{150}Sm$ (no Coulomb)& 12.5& 0.145& 6.19
\end{tabular}
\end{table}
\end{document}